\documentclass[12pt,single]{article} 
\usepackage{anysize}

\usepackage[dvips]{graphicx} 
\usepackage[hang,small,bf]{caption}
\usepackage{amsmath}
\usepackage{amssymb} 
\usepackage{amsfonts}

\begin{document}
%\vspace{-0.3cm}

\begin{center}
\section*{Muon-proton Scattering   }
\end{center}    

\subsection*{E. Borie}
\subsubsection*{Karlsruhe Institute of Technology, Association  EURATOM-FZK,  \\  
Institut f\"ur Hochleistungsimpuls and Mikrowellentechnik (IHM), \\ 
Kaiserstrasse 12, 76131 Karlsruhe,    Germany  }
%P.O.Box 3640, D-76021 Karlsruhe, Germany }                     
%Hermann-von-Helmholtzplatz~1,\\ D-76344 Eggenstein-Leopoldshafen, Germany}

%\vspace*{-.3cm}    

%\medskip
%%PACS NUmbers  36.10.-k (exotic atoms);  
%% 36.10.Dr (muonic atoms, muonium, positronium)
%%  maybe 31.30.Jv.  relativistic and QED effects in atoms+molecules
%%what are 12.20.Ds, 06.20.Jr (Pachucki96), 03.65.Pm? 
%% 12.20.Ds  QED calculations;  03.65.Pm relativistic wave equations in QM
%% 06.20.Jr determination of fundamental constants
\vspace{1.5cm}

% version 11. July, 2012
%%\bigskip
%%PACS Numbers   

\vspace{1.2cm}

\subsection*{Abstract}
\vspace{-0.2cm}
A recent proposal to measure the proton form factor by means of
muon-proton scattering will use muons which are not ultrarelativistic
(and also not nonrelativistic).  The usual equations describing the
scattering cross section use the approximation that the scattered lepton
(usually an electron) is ultrarelativistic, with $v/c$ approximately
equal to~1.  Here the cross section is calculated for all values of the
energy.  It agrees with the standard result in the appropriate limit.

%\begin{flushleft}

\subsection* { Introduction }  
\vspace{-0.2cm}
%(motivation for the proposal is the proton radius puzzle)  maybe get a
%copy of the proposal (at least its motivations)
A proposal for muon-proton scattering at PSI \cite{experiment} has been made
 in an attempt to help resolve the proton radius 
puzzle.  A measurement made on the basis of the Lamb Shift in muonic
hydrogen \cite{rpohl-experiment} disagrees with the radius measured in
atomic spectroscopy and electron scattering experiments 
\cite{codat06,mainz2010,jlab}.   The proposal will directly test whether
or not $\mu-p$ and $e-p$ scattering are the same and will perform
measurements with $\mu^{\pm}$ and $e^{\pm}$ at low $Q^2$ in order to
study the two-photon exchange contributions in greater detail.
Since the muon is about 206.7 times heavier than the electron \cite{RMP}, 
 for the energies mentioned in the proposal, the muons are 
neither ultrarelativistic nor nonrelativistic.  For the muon momenta given 
%(and in the presentation by Gilman on February 22, 2012 at PSI)  
in the proposal the value of v/c for the incoming lepton is between
0.7 and 0.9, while the standard expressions for the scattering cross section  
are valid only for v/c very close to 1.  
  The standard kinematics assumptions made in the analysis of e-p scattering 
will not all be valid in the case of an experiment on mu-p scattering.  
The cross sections have been calculated without these approximations
\cite{Landau,Preedom}, but since these results seem to have been
forgotten, it is worth presenting another calculation of the basic cross
section without them. 
%However, everything can be calculated without these approximations.
%Calculating the basic cross section without such approximations is
%straightforward, and is presented here.

% The radiative corrections when Q^2/m^2_(lepton) is of the order unity (which it will be for this experiment) will, at least for a significant part, involve rather intractable integrals.  Some of these have been given in early literature, but sorting out everything will take some time and effort.
 
According to the proposal, scattering of negative and positive muons
(and electrons) will be studied.  The muon momenta will be in the range  
(115-210)\,MeV/c with scattering angles in the range 20$^{\circ}$ 
to 100$^{\circ}$, corresponding to $Q^2$ in the range (0.01-0.1)\,(GeV/c)$^2$.
For comparison,  $m^2_{\mu}c^2$\,=\,0.01116\,(GeV/c)$^2$. 
%%  if I find Ron's reason for giving incoming lepton momenta in one of
%%  the emails, make a note of it here.

%The approximations made for the radiative corrections will also not be
%valid, since $ Q^2/m^2$ is of  order unity (between 0.9 and 9) while the   
%formulas used to calculate them assume that $Q^2/m^2 \gg 1$

\newpage
Numerical values for the muon energy ($E\,=\,\sqrt{p^2+m^2}$) and %\\
velocity \\ 
($\beta\,=\,|\vec{p}/E|\,=\,v/c$) corresponding to the incoming
momenta in the proposal are given by:
\begin{center} 
    \begin{tabular}{|ccc|}
  \hline
                  p (MeV/c)  &  E(MeV)      &    $\beta$  \\
 \hline
           115  & 156.17 &   0.7364  \\
           153  & 185.94 &   0.8229  \\
           210  & 235.08 &   0.8933  \\
   \hline
\end{tabular}        
\end{center}   
Obviously the approximation $\beta\,\approx\,1$ is not valid for the 
 energies considered in the proposal.  The radiative corrections to the 
scattering cross section are functions of $Q^2/m^2_{\mu}$, which is in
the range of approximately 0.9-9.0.  The usual 
formulas \cite{Tsai,Mo-Tsai,max-tjon}, which assume that $Q^2/m^2 \gg 1$, 
%this quantity to be either very large or very small, 
 will not be accurate.   

%In the numerical calculations the fundamental constants from  CODATA
% 2002  (\cite{codat02}) are used, i.e.:  
%$\alpha^{-1}$, $\hbar c$, $m_{\mu}$, $m_e$, $m_p$\,=\,137.0359991,
% 197.32697\,MeV$\cdot$fm,  105.658369\,MeV, \\
% 0.5109989\,MeV, 938.272 \,MeV, respectively.     

%% Kinematic variables\\
The convention of Bjorken and Drell \cite{Bjorken-Drell} will be used.  
The metric used is defined by 
\mbox{$p_i\cdot p_j\,=\,E_i E_j - \vec{p_i}\cdot \vec{p_j}$.}
 Also $m$ is the lepton rest mass, $M$ is the target rest mass, and 
$\alpha\,=\,e^2/4\pi$.  Use $p_1$ and $p_3$ for the incoming and
outgoing muon four-momenta, and $p_2$ and  $p_4$ for the incoming and
outgoing proton four-momenta, respectively.  In the lab system we have 
$p_1=(E,\vec{p})$  $p_3=(E',\vec{p'})$, $p_2=(M,0)$, $p_4=(M+\omega,\vec{q})$.  
Here $q\,=\,p_1-p_3\,=\,p_4-p_2$, and $\omega\,=\,q_0\,=\,E-E'$. 
It is useful to observe that 
\mbox{$q^2\,=\,2m^2-2p_1\cdot p_3\,=\,2M^2-2p_2\cdot p_4\,=\,-2M\omega$.}
This is simply a result of energy conservation.  Since $q^2$ is negative
with the metric used here, it is sometimes convenient to define 
$Q^2\,=\,-q^2$.
The proton current is taken to have the usual on-shell form,  characterized 
by 
\begin{equation*}
 \Gamma_{\mu}~=~F_1(q^2) \gamma_{\mu} + \kappa F_2(q^2) 
       \frac{i \sigma_{\mu \nu} q^{\nu}}{2 M} 
%\left(\frac{\alpha}{\pi}\right)^3 \,\approx \,0.00761 \, \textrm{meV} 
%%\label{eq:E-6}
\end{equation*}
Here $\kappa$ is the anomalous magnetic moment of the proton.  
The so-called Sachs form factors are related to $F_1$ and $F_2$  by
$G_M\,=\,F_1+\kappa F_2,~~G_E\,=\,F_1-\dfrac{Q^2}{4M^2}\kappa F_2$.  \\
Or, $F_1\,=\,(G_E+\dfrac{Q^2}{4M^2} G_M)/(1+\dfrac{Q^2}{4M^2})$ and 
$\kappa F_2 \,=\,(G_M -G_E)/(1+\dfrac{Q^2}{4M^2}) $

For the calculation of the matrix element, the Gordon decomposition
(\cite{Bjorken-Drell}) 
\begin{equation*}
 \bar{u}(p_4) \frac{i \sigma_{\mu \nu} q^{\nu}}{2 M}u(p_2)\,=\,
   \bar{u}(p_4)\big[\gamma_{\mu} - \frac{(p_2+p_4)_{\mu}}{2 M} \big] u(p_2)
%\left(\frac{\alpha}{\pi}\right)^3 \,\approx \,0.00761 \, \textrm{meV} 
%\label{eq:E-1}
\end{equation*}
is very useful.   As a result, one may write 
\begin{equation}
\bar{u}(p_4) \Gamma_{\mu} u(p_2)~=~\bar{u}(p_4) 
\big[(F_1+ \kappa F_2) \gamma_{\mu} - \kappa F_2 \frac{(p_2+p_4)_{\mu}}{2 M} \big] u(p_2)
\label{eq:E-2}
\end{equation}

\subsection*{Scattering  Cross Section }
\vspace{-0.2cm}
Following Chap. 7 of Ref.\,\cite{Bjorken-Drell}, the invariant matrix element 
for scattering of a charged lepton from a proton in Born approximation is 
given by
\begin{equation}
 \mathfrak{M}_{fi}~=~\bar{u}(p_3) \gamma^{\mu} u(p_1) \frac{e^2}{q^2} 
\bar{u}(p_4) \Gamma_{\mu} u(p_2)   
%\left(\frac{\alpha}{\pi}\right)^3 \,\approx \,0.00761 \, \textrm{meV} 
\label{eq:E-3}
\end{equation}
The cross section for 
scattering of the charged lepton into a given solid angle 
$d\Omega'$ about an angle $\theta$ is given by 
\begin{equation}
 \frac{d \sigma}{d \Omega'}\,=\,\frac{m^2 M}{4 \pi^2} 
 \frac{p'/p}{M+E-(pE'/p')cos\theta}  |\mathfrak{M}_{fi}|^2
%\left(\frac{\alpha}{\pi}\right)^3 \,\approx \,0.00761 \, \textrm{meV} 
\label{eq:E-4}
\end{equation}
In Eq.\,\ref{eq:E-4} it is assumed that $ |\mathfrak{M}_{fi}|^2$ has  
been averaged over initial spins and summed over final spins.
%The traces involved will be given in an Appendix.  
The final result is given by 
\begin{equation}  \begin{split}
 \frac{d \sigma}{d \Omega'} & \,=\,\frac{\alpha^2}{q^4} 
 \frac{p'/p}{1+(E-(pE'/p')\cos\theta)/M} 
  \Big[ G^2_E \frac{(4EE' + q^2)}{1-q^2/4M^2}  \\ 
 & ~~~~+ G^2_M \Big((4EE'+ q^2)\big(1-\frac{1}{1-q^2/4M^2}\big)+\frac{q^4}{2M^2} 
  + \frac{q^2 m^2}{M^2} \Big) \Big]
%\left(\frac{\alpha}{\pi}\right)^3 \,\approx \,0.00761 \, \textrm{meV} 
\end{split}
\label{eq:E-5}
\end{equation}
Recall that generally $-q^2\,=\,Q^2=2M(E-E')\,=\,2(EE' - pp'\cos\theta - m^2)$

If the limit of very high lepton energies, one has  
\begin{equation*}
  p \approx E, ~~~ p' \approx E', ~~~ q^2 \approx -4EE' \sin^2(\theta /2) 
\end{equation*}
In this case, the cross section given in Eq.\,\ref{eq:E-5} reduces to
\begin{equation} \begin{split}
 \frac{d \sigma}{d \Omega'} & \,=\,\frac{\alpha^2 \cos^2(\theta /2)}
 {4E^2\sin^4(\theta/2)} \frac{1}{1+ 2E \sin^2(\theta/2)/M}  \\
 & ~~~~\Big[\frac{Q^2}{2M^2} G^2_M \Big(\frac{1}{1+Q^2/4M^2}+ 2\tan^2(\theta /2) \Big)
    +  \frac{G^2_E}{1+Q^2/4M^2}\Big]
%\left(\frac{\alpha}{\pi}\right)^3 \,\approx \,0.00761 \, \textrm{meV} 
\end{split}
\label{eq:E-6}
\end{equation}
This agrees with the expression given in Sec.4 of ref.\,\cite{karshenboim} 
(and with equivalent expressions in other work). 

%Here $\mathfrak{M}_{fi}$ is the invariant matrix element as defined in
%Ref.\,\cite{Bjorken-Drell} and in Eq.\,\ref{eq:E-3} it is assumed to
%have been averaged over initial spins and summed over final spins.

%% example of split equations 
%For deuterium, with $s_2=1$, the corresponding hyperfine splitting is 
%\begin{equation*}  \begin{split}
%\Delta  E_{ns} &\,=\,  \frac{2 (\alpha Z)^4 m_r^3}{3 n^3 m_{\mu} m_D}
% \cdot (1 + \kappa_D) \cdot (1 + a_{\mu}) \cdot [F(F+1) - 11/4] \\
%  & \,=\, \frac{\beta_D}{2} \cdot (1 + a_{\mu}) \cdot [F(F+1) - 11/4] 
%   \,=\, (8/n^3) \cdot 2.04766\,meV \cdot [ \delta_{F,3/2} - 2  \delta_{F,1/2}]
%\end{split}
%\end{equation*}
%\noindent for a total splitting of 6.14298\,meV in muonic deuterium.  

%figure examples
\begin{figure}[!h]
%\centering\includegraphics[width=0.80\linewidth]{Comparison-Picture2.eps}
%%%%  fig3.ps is color, fig4.ps is black+white 
%\centering\includegraphics[width=0.60\linewidth,angle=270]{vp/scatt/fig3.ps}
%\centering\includegraphics[width=0.60\linewidth,angle=270]{trento/fig3.ps}
centering\includegraphics[width=0.60\linewidth,angle=270]{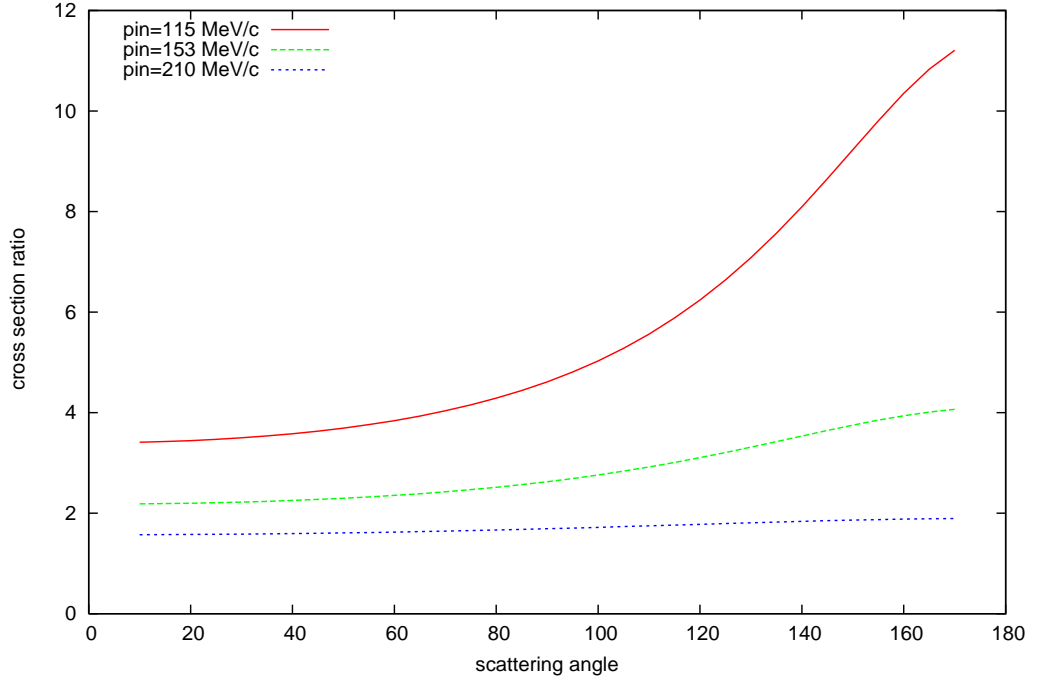}
\caption {Ratio of the exact scattering cross section to the cross
  section calculated with the commonly used relativistic approximation. }
%  solid curve: p_{in}=115\,MeV/c,~ long dashed: p_{in}=153\,MeV/c,~
%  short dashed: p_{in}=210\,MeV/c }
% red curve: p_{in}=115\,MeV/c;~green curve: p_{in}=153\,MeV/c;~blue curve: p_{in}=210\,MeV/c }
 \label{fig1}
\end{figure}
 Figure\,\ref{fig1} shows  the ratio of the cross section calculated
 with the exact formula (Eq.\,\ref{eq:E-5}) to that calculated with 
 Eq.\,\ref{eq:E-6}, but with exact muon kinematics.  The form factors
 were taken from a parametrization by \cite{kelly}.  The difference is
 significant, especially at lower values of incident momentum.

\newpage
\subsection*{Radiative Corrections }
\vspace{-0.2cm}

Radiative corrections to the electromagnetic properties of a muon or an
electron modify the lepton scattering cross sections (and also produce
energy level shifts in atoms).

According to the standard theory, the cross section for scattering is
altered by a factor $1-\delta$ where, for potential scattering at high
momentum transfer 
\cite{Schwinger},  
\begin{equation*}
\delta \approx \frac{2 \alpha}{\pi} \big[(\ln(Q^2/m^2)-1)\ln(E/\Delta E) \big]
  + \delta_{vertex}  + \delta_{VP} 
%\label{eq:E-2}
\end{equation*}
Here $\Delta E $ is the energy acceptance of the detector, and is
related to the fact that the lepton can emit very low-energy photons
that would not be detected.  The cross section for this process (soft
photon bremsstrahlung) is infrared divergent; the divergence is canceled
by a similar infrared divergence in the vertex correction.  Since then, the
 calculations have been extended by many authors \cite{Tsai,Mo-Tsai,max-tjon} 
to include target recoil, radiation from the target, the target vertex 
correction, and two-photon exchange. 

 The energy of the recoiling proton will be rather small ($< (50-60)$\,MeV) 
in this experiment, so that soft photon radiation from the proton will
contribute very little to the radiative corrections.  The contribution
from two photon exchange is beyond the scope of this paper. 
%(see a couple of other references???).  
For this reason, only the purely leptonic contributions to the radiative 
correction will be considered.  In most of the 
literature \cite{Tsai,Mo-Tsai,max-tjon}, it has been assumed that $Q^2/m^2$ 
is very large.  It turns out that when this is not the case
(as in the proposed experiment), there will be an extra contribution to
the radiatively corrected cross section arising from the fact that the
lepton vertex correction gives rise to \textbf{two} form factors $F_1$
and $F_2$ ($F_2$ corresponds to the anomalous magnetic moment in the
limit $Q^2 \rightarrow 0$).  This does not simply multiply the
uncorrected cross section.  Exact expressions for these form factors
have been given by Feynman \cite{Feynman}, in two textbooks 
\cite{Landau,akhiezer}, and more recently by Barbieri et al. \cite{barbieri}.
They all agree with each other, except for different notation.  Here the
notation used in Ref.\,\cite{barbieri} will be used.

To lowest order in $\alpha$ the form factors are given by 
\begin{equation*}
  F_1(t)\,\approx 1 + \frac{\alpha}{\pi} F^{(2)}_1 (t)
\end{equation*}
\begin{equation*}
  F_2(t)\,\approx \frac{\alpha}{\pi} F^{(2)}_2 (t)
\end{equation*}
As usual, $t=q^2=(p_1-p_3)^2$.  \\

For spacelike $q^2=-Q^2$ the form factors can be expressed in terms of a
variable $\Theta$ that is related to the momentum transfer by 
$\sinh(\phi) = Q/(2 m)$ and    
\begin{equation*}%      
% \Theta \,=\, \frac{\coth(\phi) - 1}{\coth(\phi) +1} 
 \Theta \,=\, \frac{1 - \tanh(\phi)}{1+\tanh(\phi)} 
%\label{eq:f-phi}  
\end{equation*}    
It is useful to note that $\ln(\Theta)\,=\,-2\phi$ and that 
$Q^2/m^2\,=\,(1-\Theta)^2/\Theta\,=\,4\sinh^2(\phi)$. \\
In terms of this variable, 
\begin{equation}  \begin{split}
  F_1(t)-1\,=\,& \frac{\alpha}{\pi} \Big[\ln\big(\frac{m}{\lambda}\big)
  \Big(1+\frac{1 + \Theta^2}{1-\Theta^2} \ln(\Theta)\Big)  \\
  & -1 - \frac{3 \Theta^2+2\Theta+3}{4(1-\Theta^2)} \ln(\Theta)  \\
  & + \frac{1 + \Theta^2}{1-\Theta^2}\Big(\frac{\pi^2}{12} +  L_2(-\Theta) 
    - \frac{1}{4} \ln^2(\Theta) + \ln(\Theta)\ln(1+\Theta) \Big)   \Big]
\end{split}
\label{eq:F1}
\end{equation}
(The infrared divergent part will be compensated by contributions from
 bremsstrahlung of soft photons.)  \\
\noindent and
\begin{equation*}
  F_2(t)\,=\, - \frac{\alpha}{\pi} \frac{\Theta}{1-\Theta^2} \ln(\Theta)
\end{equation*}
%\begin{equation*}   (also)
%  F_2(t)\,=\, - \dfrac{\alpha}{\pi} \frac{\phi}{sinh(2\phi)}
%\end{equation*}
%This result has also been given by other authors \cite{Landau,Feynman}.
%In the limit of vanishing $Q^2$, the result is well known: 
%$F_2(0)\,=\, \dfrac{\alpha}{2\pi}$.  \\
In the limit of very small $Q^2$, 
$F_2(0)\,\rightarrow\,\dfrac{\alpha}{\pi}\big(\dfrac{1}{2}-\dfrac{Q^2}{12
  m^2}\big)$.  \\
The value of $F_2(0)$ is well known.
At very high momentum transfers,  $F_2(t)$ becomes negligibly small:  
\begin{equation*}   
  F_2(t) \,\rightarrow \, -\frac{\alpha}{\pi} \frac{m^2}{Q^2}\ln(m^2/Q^2)
\end{equation*}
% (it is proportional to $m^2/Q^2 \ln(Q^2/m^2) $
%%%%%%% changes October 20:

In the limit of high momentum transfer, $F_1(t)$ becomes 
\begin{equation}   
  F_1(t)-1 \,\rightarrow \, \frac{\alpha}{\pi}\Big[\big(1-\ln(Q^2/m^2)\big)
  \ln(m/\lambda) -1 + \frac{3}{4}\ln(Q^2/m^2)
     + \frac{\pi^2}{12}  - \frac{1}{4} \ln^2(Q^2/m^2)  \Big]
\label{eq:vertex}  
\end{equation}
For the momentum transfers of interest in this experiment,  $F_2(t)$ is
comparable to the other radiative corrections, and it results in a
nonmultiplicative contribution to the radiative corrections.  A 
calculation of this  contribution to the radiatively corrected cross
section for lepton-proton scattering will be given further below.
%\smallskip
The muon current including (vertex) radiative corrections of order $\alpha$ will
have the form 
\begin{equation*}
\bar{u}(p_3) \big[(F_1+ F_2) \gamma_{\mu} -F_2 \frac{(p_1+p_3)_{\mu}}{2 m} \big]u(p_1)
%\label{eq:E-2}
\end{equation*}
where the Gordon decomposition has been used.
%Here (see ref.\cite{barbieri})
%\begin{equation*}
%  F_1(t)\,\approx 1 + \frac{\alpha}{\pi} F^{(2)}_1 (t)
%\end{equation*}
%\begin{equation*}
%  F_2(t)\,\approx \frac{\alpha}{\pi} F^{(2)}_2 (t)
%\end{equation*}
%As usual, $t=q^2=(p_1-p_3)^2$.  \\
%%  also some changes here for the slides, and preprint Oct 21, 2012
The radiative corrections also include contributions from vacuum
polarization, so that in the matrix element for scattering the
contributions from muon and electron loops must be added to the term
proportional to $\gamma_{\mu}$.
These are:
\begin{equation*}
  \frac{\alpha}{3 \pi} \Big[\frac{1}{3} +  (\coth^2(\phi)-3) 
  (1 - \phi \cdot \coth(\phi))\Big]  \,=\, \frac{\alpha}{\pi} U_{2m} 
\end{equation*}
\noindent for muon loops and
\begin{equation*}
 \frac{ \alpha}{3 \pi} \Big[\ln(\frac{Q^2}{m_e^2}) - \frac{5}{3} \Big] 
 \,=\, \frac{\alpha}{\pi} U_{2e} 
\end{equation*}
%\begin{equation*}
% \delta_{\mu VP} \,=\, \frac{2 \alpha}{3 \pi} \Big[\frac{1}{3} +
% (\coth^2(\phi)-3)   (1 - \phi \cdot \coth(\phi))\Big]  
%\end{equation*}
%\noindent and
%\begin{equation*}
% \delta_{eVP} \,=\, \frac{2 \alpha}{3 \pi}  \Big[\ln(\frac{Q^2}{m_e^2})
% - \frac{5}{3} \Big]
%\end{equation*}
for electron loops.  Note that for the contribution due to electron loops, 
one may use the usual high momentum transfer approximation.  
%The vertex contribution is simply  $\delta_{vertex}\,=\,2(F_1+F_2-1)$

The lepton trace will be evaluated to leading order in $\alpha/\pi$, so
that $F_2^2\,\approx\,0$ and
\begin{equation*}
 (F_1 + F_2)^2\,\approx 1+ \frac{2\alpha}{\pi}( F^{(2)}_1+ F^{(2)}_2),~~~~~~~~~~~~F_1
 F_2\,\approx  \frac {\alpha}{\pi} F^{(2)}_2 
\end{equation*}
Inclusion of the vacuum polarization contributions replaces  
$F^{(2)}_1+ F^{(2)}_2$  \\ by $F^{(2)}_1+ F^{(2)}_2 +  U_{2m} +  U_{2e} $. 

A possible contribution to the vacuum polarization  from hadronic loops
has not been included.  The earliest calculations of this contribution to 
the energy levels of muonic atoms \cite{hadron1,hadron2}  have indicated  
that, at least in the limit of small $Q^2$, this contribution is approximately 
0.66 times the contribution from muon loops.  Generally, it would be equal to 
$(2\alpha/\pi) \Pi_H(Q^2)$, where $\Pi_H(Q^2)$ is defined in 
Ref.\,\cite{hadron1}.  Since these papers were written, much better data 
for the cross section for $e^+ e^- \rightarrow $\,hadrons have become
available, so that a recalculation of $\Pi_H(Q^2)$ as a function of $Q^2$ 
would be desirable.  The same cross section enters into the correction 
to the anomalous magnetic moment of the muon.  

%The  vacuum polarization contribution to the
%radiative correction $\delta_{VP}$ is the sum of contributions from muon
%and electron loops, with 

One can show that the infrared divergent contribution to the real
soft-photon bremsstrahlung cross section $d \sigma_b$ given in Eq.(4.7)
of Ref.\,\cite{max-tjon} (the coefficient of  $\ln(2\Delta E/\lambda)$) 
is 
\begin{equation*}
-\frac{2 \alpha}{\pi}\,\ln(2\Delta E/\lambda)\big[1-2\phi \coth(2\phi)\big]
\end{equation*}
where  $d \sigma_0$ is the uncorrected cross section and $\Delta E$ is
the energy resolution of the detector.  Since \\
$1+\frac{1+\Theta^2}{1-\Theta^2}\ln(\Theta)=1-2\phi \coth(2\phi)$
one can combine this with the infrared divergent part of Eq.\,\ref{eq:F1}   
to obtain a contribution $\delta_{rad}$ to the radiative correction: 
\begin{equation*}
\delta_{rad}=-\frac{2 \alpha}{\pi}\,\big[1-2\phi \coth(2\phi)\big] \ln(2\Delta E/m)
\end{equation*}
Other noninfrared divergent contributions will also contribute; in the
limit of large values of $Q^2/m^2$, these were given in Eq.(4.14) of
Ref.\,\cite{max-tjon} and it can be seen that the contribution
proportional to $\ln^2(Q^2/m^2)$ in Eq.\,\ref{eq:vertex} is also
canceled by some of these terms.  This will probably happen also for the
present case.    
%It will be shown below that the contribution from the vertex correction
%to the total radiative 
%correction $\delta$ is $\dfrac{2\alpha}{\pi}( F^{(2)}_1+ F^{(2)}_2)$.   
% If this contribution (for large values of $Q^2/m^2$) is
% added to the contribution from real  
%soft-photon bremsstrahlung given in Eq.(4.14) of Ref.\,\cite{max-tjon}, one
% obtains for the contribution to $\delta$ 
%\begin{equation}  \begin{split}
%  \delta_{IR}+ \delta_{vertex} \,\simeq\,- \frac{2 \alpha}{\pi} & 
% \Big[\ln\big(\frac{\eta\Delta E}{\sqrt{E E'}}\big) 
% \Big(1+\frac{1 + \Theta^2}{1-\Theta^2} \ln(\Theta)\Big)  \\ 
%  &  + \frac{3}{4}\ln(Q^2/m^2) -1 - \frac{\pi^2}{12} -
%  \frac{1}{4} \ln^2(\eta) +  \frac{1}{2}L_2(\cos^2(\theta/2))  \Big]
%\end{split}
%\end{equation}
%where $\eta\approx E/E'$.  
%It should be noticed that in addition to the terms involving ln(m/$\lambda$),  
%the contribution proportional to $ \ln^2(Q^2/m^2)$ in Eq.\,\ref{eq:vertex}  
% is also canceled.  
%% reformulate below:
%

%The exact expression for the vacuum polarization contribution to the
%radiative correction $\delta$ is
%\begin{equation*}
% \delta_{VP} \,=\, -\frac{ \alpha}{3 \pi} \cdot  F(\phi) 
%\end{equation*}
%\noindent where  
%\begin{equation*}      
% F(\phi) \,=\, \frac{1}{3} + (\coth^2(\phi)-3) \cdot 
%  [1 - \phi \cdot \coth(\phi)]
%\label{eq:f-phi}  
%\end{equation*}    

%%%%  here the nonmultiplicative contribution to the radiative
%%%%  correction from  $F_2(t)$.

%\newpage
 The  additional contribution to the radiatively corrected cross section
 due to the  presence of $F_2(Q^2)$ will now be given.  For this  
 %For the calculation of the additional contribution to the radiatively
 %corrected cross section, 
the revised lepton trace (compare with Eq.\ref{eq:lept}) is  needed.  
It is given by
%\begin{equation*}\begin{split}
% \frac{(F_1 + F_2)^2}{m^2}  &  [p_3^{\mu} p_1^{\nu} +  p_1^{\mu}  p_3^{\nu}
% + g^{\mu \nu} q^2/2 ]  \\  &  - \frac{F_1 F_2}{m^2} (p_1+p_3)^{\mu} (p_1+p_3)^{\nu}
%\end{split}
%\end{equation*}
\begin{equation*}
 \frac{(F_1 + F_2)^2}{m^2} [p_3^{\mu} p_1^{\nu} +  p_1^{\mu}  p_3^{\nu} + g^{\mu \nu} q^2/2 ]  
   - \frac{F_1 F_2}{m^2} (p_1+p_3)^{\mu} (p_1+p_3)^{\nu}
\end{equation*}
The term with $ (F_1 + F_2)^2\,\approx 1$ corresponds to the uncorrected
cross section and the term of order  $\alpha/\pi$ will contribute a
factor multiplying the uncorrected cross section, thus giving 
$\delta_{el}\,=\,\delta_{vertex} + \delta_{VP}$.  Here 
 $\delta_{vertex}\,=\,2(F_1+F_2-1)$  
(Since the infrared divergent part of $F_1$ will be canceled by contributions
from radiation of real soft photons by the muon,  it is not included in
any of the numerical examples.)  
%The vertex contribution is simply  $\delta_{vertex}\,=\,2(F_1+F_2-1)$
and  $\delta_{VP}\,=\,(2 \alpha/\pi)(U_{2m} + U_{2e})$. 
 For the limiting case of very high values of $Q^2/m^2$, the expression
 for  $\delta_{el}$ agrees with the corresponding expression in
 Eq.(3.36) of Ref.\,\cite{max-tjon}
The contributions to  $\delta_{el}$ from the vertex
correction and from vacuum polarization (without the contribution from
radiation of real (soft) photons) were also given (with a different, but
equivalent notation) in Chap.\,51 of Ref.\,\cite{akhiezer}.  
The presence of an additional correction, not proportional to the
uncorrected cross section, was also mentioned  in that work, but only in
connection with scattering from a Coulomb potential. Hence only the 
contribution of  the last term in the lepton trace (proportional to $
(p_1+p_3)^{\mu} (p_1+p_3)^{\nu}$) will have to be calculated for the
case of muon-proton scattering. 

%Before doing this, the contributions to $\delta$ from the vertex
%correction and from vacuum polarization (without the contribution from
%radiation of real (soft) photons) will be given, for completeness.
%They were also given (with a different, but equivalent notation) in
%Chap.\,51 of Ref.\,\cite{akhiezer}.  The presence of an additional
%correction, not proportional to the uncorrected cross section, was also
%mentioned  in that work.  

%Here the contribution from the emission of
%real soft photons is not included. 
%  also the sign is for the convention multiply equation \ref{eq:

%The  vacuum polarization contribution to the
%radiative correction $\delta_{VP}$ is the sum of contributions from muon
%and electron loops, with 
%\begin{equation*}
% \delta_{\mu VP} \,=\, \frac{2 \alpha}{3 \pi} \Big[\frac{1}{3} + (\coth^2(\phi)-3) 
%  (1 - \phi \cdot \coth(\phi))\Big]  
%\end{equation*}
%\noindent and
%\begin{equation*}
% \delta_{eVP} \,=\, \frac{2 \alpha}{3 \pi}  \Big[\ln(\frac{Q^2}{m_e^2})
% - \frac{5}{3} \Big]
%\end{equation*}
%Note that for the contribution due to electron loops, one may use the usual high momentum transfer approximation.  

%%%%%%  end of changes Oct 20, 22012
The additional term (which will be multiplied by $F_2$) will now be
calculated.   
The proton trace is evaluated in the  Appendix and is equal to
%The proton trace was evaluated in the first Appendix and is equal to
\begin{equation*}  \begin{split}
  \frac{G^2_M}{M^2} & [p_{2\mu}p_{4\nu} + p_{2\nu}p_{4\mu} + g^{\mu \nu} q^2/2] \\
 & + \frac{G_E^2-G_M^2}{2 M^2 (1-q^2/4M^2)}  (p_2+p_4)_{\mu} (p_2+p_4)_{\nu} 
 \frac{1}{M^2}(p_2 \cdot p_4+M^2) 
\end{split}
\end{equation*}
The additional term in the spin-averaged square of the matrix element
$|\mathfrak{M}_{fi}|^2$ %(Eq.\,\ref{eq:matel}) 
is then (without the factor $ -F_2 $)
%(without the factor $-(\alpha/\pi) F_2 $)
\begin{equation*}  \begin{split} 
% \hspace{-0.5cm} 
\frac{4 \pi^2 \alpha^2}{m^2 M^2(q^2)^2}     
& \Big[ G^2_M  \big[2p_2\cdot(p_1+p_3) p_4\cdot(p_1+p_3)+ (p_1+p_3)^2 q^2/2\big]  \\ 
& ~~ + \frac{G_E^2-G_M^2}{2(1-q^2/4M^2)}\big((p_1+p_3)\cdot(p_2+p_4)\big)^2 \Big]  
\end{split}
%\label{eq:matel}
\end{equation*}

Evaluation of the scalar products gives finally 
\begin{equation} 
% \begin{split}
  \frac{4 \pi^2 \alpha^2}{m^2 (q^2)^2}    
 \Big[ G^2_M \big[2(E+E')^2 +  (4m^2-q^2)\frac{q^2}{2M^2}\big]  
 + 2\frac{G_E^2-G_M^2}{(1-q^2/4M^2)}(E+E')^2 \Big]  
%\end{split}
\label{eq:extra}
\end{equation}
As in Eq.\,\ref{eq:E-4}, to obtain the contribution to the radiatively
corrected cross section, this will be multiplied by  
$\dfrac{m^2 }{4 \pi^2}  \dfrac{p'/p}{1+(E-(pE'/p')\cos\theta)/M}$ to obtain
%$\dfrac{m^2 M^2}{4 \pi^2}  \dfrac{p'/p}{M+E-pE'/p'cos\theta}$ to obtain
the cross section corresponding to the additional term.  This will be
denoted by $d \sigma_1$ while the cross section given in
Eq.\,\ref{eq:E-5} will be denoted by $d \sigma_0$

\begin{figure}[!h]
%\centering\includegraphics[width=0.80\linewidth]{Comparison-Picture2.eps}
%%%%  fig2.ps is for radiative corrections- corrected in fig4.ps  
%%% for home: leave out figure, or substitute fig3.ps since the figure
%% is not available
%\centering\includegraphics[width=0.60\linewidth,angle=270]{trento/fig4.ps}
%\centering\includegraphics[width=0.60\linewidth,angle=270]{vp/scatt/fig4.ps}
\centering\includegraphics[width=0.60\linewidth,angle=270]{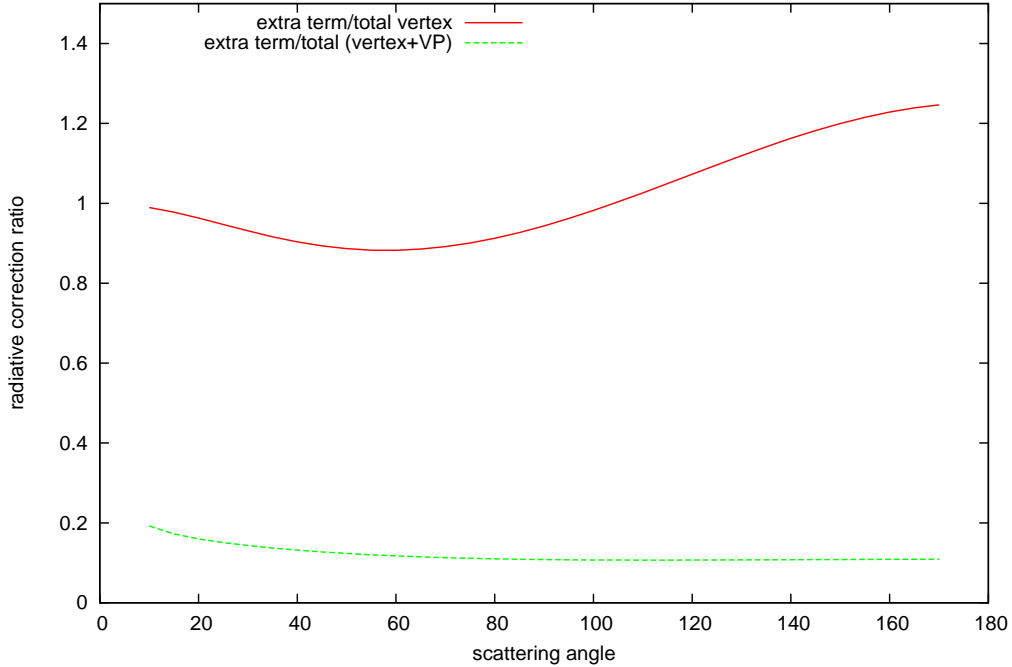}
\caption {Ratio of the additional contribution to the radiatively
  corrected cross section  to contributions to the  usual radiatively
  corrected cross section} 
%  solid curve: p_{in}=115\,MeV/c,~ long dashed: p_{in}=153\,MeV/c,~
%  short dashed: p_{in}=210\,MeV/c }
% red curve: p_{in}=115\,MeV/c;~green curve: p_{in}=153\,MeV/c;~blue curve: p_{in}=210\,MeV/c 
 \label{fig2}
\end{figure}

 Figure\,\ref{fig2} shows  the ratio of the additional contribution to
the radiatively corrected cross section ($F_2 \times d \sigma_1$)
%(Eq.\,\ref{eq:extra})) 
to the usual contribution to the radiatively corrected cross section from the
noninfrared part of the vertex correction 
%($F_1-1+F_2)\times$(Eq.\,\ref{eq:E-5}) (red curve) for an initial
($2(F_1-1+F_2)\times d \sigma_0$)
%(Eq.\,\ref{eq:matel}) 
(solid (red) curve) for an initial muon momentum of 153\,MeV/c.  
The dashed (green) curve shows a similar ratio, but including 
the contribution due to vacuum polarization (muon and electron loops).
The contribution from the electron loops is significantly larger than
the other contributions.  For the momentum transfers of interest, the
nonmultiplicative contribution to the radiative corrections is not negligible.  
However, the calculation of the additional correction is straightforward.  

 %%%%  must check the relative sign for VP and vertex (also muVP vs eVP)
%%%% also maybe a factor 2 in the denominator for the red curve 
% (2(F_1+F_2) for the multiplicative vertex)-but need the correct
% factors relative to VP before recalculating (these as in Akhiezer eq 51.13)

%%%%% what to do if the extra term is to be included in a multiplicative delta?
%%% Unlike the case of potential scattering, if all terms proportional to
%%% $F_2$ are included in $\delta$ then this correction depends in a rather
%%% complicated manner on the nucleon form factors.  Experimentalists may  
%%% prefer to treat it separately in their analysis.
%\begin{equation*}
%  F_2\,=\, - \frac{\alpha}{\pi} \frac{\Theta}{1-\Theta^2} \ln(\Theta)
%\end{equation*}

It is possible to include the contribution of $F_2$ in a multiplicative
contribution to $\delta$, but this would depend on the nucleon form factors
 in a rather complicated manner.  In this case   
%$\delta_{el}\,=\,\delta_{vertex} + \delta_{VP}$.  Here 
 $\delta_{vertex}\,=\,2(F_1+F_2-1)$  would be replaced by   
\begin{equation*}  \begin{split} 
 \delta_{vertex}\,=\, 2(F_1-1) - & 2F_2\, Q^2 \frac{1-Q^2/4M^2}{1+Q^2/4M^2}
 \big(G^2_E + \frac{Q^2}{4M^2}G^2_M\big) \\ 
 & \cdot \Big[(4EE'-Q^2)\big(G^2_M + \frac{G_E^2-G_M^2}{(1+Q^2/4M^2)}\big)
  +   \frac{Q^2}{2M^2}(Q^2-2m^2)G^2_M \Big]^{-1}
\end{split} 
\end{equation*}
in order to include the additional term in $\delta_{el}$ rather than
treating it separately.  

\newpage
\noindent The table below gives the relative magnitude of the different contributions
for a few cases.  Recall that by definition, the contributions to lowest
order in $\alpha$ are: \mbox{$F_1-1=(\alpha/\pi)F^{(2)}_1$,}    
$F_2= (\alpha/\pi)F^{(2)}_2$   
and $\delta_{VP}\,=\,(2\alpha/\pi)(U_{2m} +  U_{2e}) $. 
Only the noninfrared part of $F^{(2)}_1$ is given in this table.   
 The vacuum polarization contribution from electron loops 
is significantly larger than the other contributions.  
To check whether higher order electron vacuum polarization
(corresponding to two loops) might be important, the value of
$(\alpha/\pi) U^2_{2e}$ was also calculated.  %\\
%%was calculated for a few cases.  \\
%values of  $F_1-1$, $F_2$,  $U_{2m}$ and  $U_{2e}$ (all without the
%overall factor $(\alpha/\pi)$), as well as $(\alpha/\pi) U^2_{2e}$.  
%\begin{center} 
%    \begin{tabular}{|rccccccc|}
% \hline
%  $\theta (^{\circ})$  &   $Q^2(GeV/c)^2$  &  $F^{(2)}_1$ & $F^{(2)}_2$ 
% & $U_{2m}$   & $U_{2e}$  & $(\alpha/\pi) U^2_{2e}$  &  $(\alpha/\pi) U_{4e}$ \\
% \hline
%  30 &  0.00611  & 0.0564 &  0.4588 &  0.0345 &  2.7582 &  0.0128 & 0.0266  \\
%  60 &  0.02133  & 0.1256 &  0.3841 &  0.1069 &  3.2146 &  0.0240 & 0.0337  \\
%  90 &  0.03907  & 0.1394 &  0.3259 &  0.1746 &  3.4164 &  0.0271 & 0.0377  \\
% 120 &  0.05394  & 0.1240 &  0.2908 &  0.2221 &  3.5239 &  0.0288 & 0.0401  \\
%   \hline
%\end{tabular}        
%\end{center}   
\begin{center} 
    \begin{tabular}{|rcccccc|}
 \hline
  $\theta (^{\circ})$  &   $Q^2(GeV/c)^2$  &  $F^{(2)}_1$ & $F^{(2)}_2$ 
 & $U_{2m}$   & $U_{2e}$    &  $(\alpha/\pi) U_{4e}$ \\
 \hline
  30 &  0.00611  & 0.0564 &  0.4588 &  0.0345 &  2.7582  & 0.0266  \\
  60 &  0.02133  & 0.1256 &  0.3841 &  0.1069 &  3.2146  & 0.0337  \\
  90 &  0.03907  & 0.1394 &  0.3259 &  0.1746 &  3.4164  & 0.0377  \\
 120 &  0.05394  & 0.1240 &  0.2908 &  0.2221 &  3.5239  & 0.0401  \\
   \hline
\end{tabular}        
\end{center}   
 It turned out to be comparable to the vertex contributions.  This 
indicates that the contribution from fourth order electron vacuum polarization 
might be comparable to other contributions.  The expression for the complete
contribution to this order has been given in Eq.\,(140) of Ref.\,\cite{RMP}
(see also Ref.\,\cite{barbieri73}).  In the limit $Q^2/m_e^2 \gg 1$, this
results in an additional contribution to $\delta_{eVP}$ of
 $2(\alpha/\pi)^2 U_{4e}$ where  
\begin{equation*}
 U_{4e} \,=\, U^2_{2e}  +  U_{2e} U_{2m}
-\frac{5}{24} + \zeta(3) +\frac{1}{4} \ln(\frac{Q^2}{m_e^2})
\end{equation*}
To compare this with the other contributions, the last column gives  
also $(\alpha/\pi) U_{4e}$.  
%This is probably the only higher order contribution that needs
%to be included.   
%It might also be numerically significant to multiply the noninfrared
%contribution to  $\delta_{vertex}$ by $1 + (\alpha/\pi)U_{2e} $, since
%$U_{2e}$ is  logarithmically enhanced.  
Other higher order contributions are probably negligible. 

The radiative corrections calculated for this example (and hence for the
proposed experiment) are somewhat smaller than the radiative corrections
encountered in electron scattering at higher momentum transfer.  For an
incoming muon momentum of 153\,MeV/c and scattering angle of $60^{\circ}$
the lowest order electron vacuum polarization contributes 1.49\% to the correction;  
including the noninfrared part of the vertex correction,
vacuum polarization due to muon loops, and fourth order electron vacuum
polarization gives a leptonic contribution to $\delta_{el}$ of 1.80\%.  
The correction varies from 1.57\% to 1.93\% as the scattering angle
varies from  $30^{\circ}$ to  $120^{\circ}$.

The extra contribution to the radiatively corrected cross section 
($-F_2 \times d \sigma_1/d\Omega$) for this example was 
$-0.000078 \times 10^{-24} cm^2/$steradian  while the   
%(Eq.\,\ref{eq:extra})) 
usual contribution to the radiatively corrected cross section from the
noninfrared part of the vertex correction 
($\delta_{el}\times d \sigma_0/d\Omega$) is $-0.000671 \times 10^{-24}cm^2/$steradian.  
The extra contribution is about 11\% of the standard contribution.

Of course, the contribution from soft photon radiation, as well as from
the proton vertex correction, and two-photon corrections should also be
included.  Since these depend on the proton structure, they should
probably be given separately.  

In conclusion, the radiative corrections for the experiment proposed in 
Ref.\,\cite{experiment} can be expected to be smaller (only a few percent)  
than is the case for electron scattering at higher momentum transfers.
They are dominated by vacuum polarization with electron loops.  The
fourth order electron vacuum polarization is comparable to the second
order muon vacuum polarization and vertex corrections, and should not be
ignored.  There is an extra, nonmultiplicative contribution to the
radiative corrections.  It can be as large as 15\% of the standard
(multiplicative) radiative correction.

%\newpage
\subsubsection*{ Acknowledgments }
\vspace{-0.2cm}
The author wishes to thank R. Gilman for extensive 
email correspondence regarding this work.

\small
\def\refname{{\normalsize References}}
%\begin{thebibliography}{99}

%\newpage
\newpage
\subsection*{Appendix }
\vspace{-0.2cm}
Here a few details of the calculation of the average over initial spins
and sum over final spins of  $ |\mathfrak{M}_{fi}|^2$ are given.  
$\mathfrak{M}_{fi}$ is given by Eq.\,\ref{eq:E-3}.  

\smallskip

\noindent  Following Chap. 7 of Ref.\,\cite{Bjorken-Drell} the spin sum and 
average of  $ |\mathfrak{M}_{fi}|^2$ is given by  
\begin{equation}  \begin{split}
 |\mathfrak{M}_{fi}|^2 & ~=~ \frac{e^4}{4(q^2)^2} \cdot
  Tr \Big[\big(\frac{\not\!p_3+m}{2m}\big)\gamma^{\mu}
  \big(\frac{\not\!p_1+m}{2m}\big)\gamma^{\nu}\Big] \cdot   \\  
 & ~~~~~~~~ Tr \Big[\big(\frac{\not\!p_4+M}{2M}\big)  \Gamma_{\mu}
  \big(\frac{\not\!p_2+M}{2M}\big)\Gamma_{\nu}\Big] 
\end{split}
\end{equation}

\noindent  The lepton trace is 
\begin{equation}
\frac{1}{m^2} [p_3^{\mu} p_1^{\nu} +  p_1^{\mu}  p_3^{\nu} + g^{\mu \nu} q^2/2 ]  
\label{eq:lept}
\end{equation}

\noindent The trace for the proton is
\begin{equation}  \begin{split}
   G^2_M & Tr \Big[\big(\frac{\not\!p_4+M}{2M}\big) \gamma_{\mu}
  \big(\frac{\not\!p_2+M}{2M}\big)\gamma_{\nu}\Big] 
  + \big(\frac{\kappa F_2}{2 M}\big)^2 (p_2+p_4)_{\mu} (p_2+p_4)_{\nu} 
 Tr \big(\frac{\not\!p_4+M}{2M}\big)\big(\frac{\not\!p_2+M}{2M}\big)\\
 &  - \frac{G_M  \kappa F_2}{2 M} \Big[(p_2+p_4)_{\mu} 
 Tr \big(\frac{\not\!p_4+M}{2M}\big)\big(\frac{\not\!p_2+M}{2M}\big)\gamma_{\nu}      
 + (p_2+p_4)_{\nu}
  Tr \big(\frac{\not\!p_4+M}{2M}\big)\gamma_{\mu}\big(\frac{\not\!p_2+M}{2M}\big)       
\Big] 
\end{split}
\end{equation}

\noindent  where Eq.\,\ref{eq:E-2} has been used. 

Evaluation of the traces gives
\begin{equation}  \begin{split}
  \frac{G^2_M}{M^2} & [p_{2\mu}p_{4\nu} + p_{2\nu}p_{4\mu} + g^{\mu \nu} q^2/2]
  + \big(\frac{\kappa F_2}{2 M}\big)^2 (p_2+p_4)_{\mu} (p_2+p_4)_{\nu} 
 \frac{1}{M^2}(p_2 \cdot p_4+M^2) \\
  &  - \frac{G_M  \kappa F_2}{M^2} (p_2+p_4)_{\mu} (p_2+p_4)_{\nu}
\end{split}
\end{equation}
The terms involving $\kappa F_2 \,=\,(G_M -G_E)/(1-\dfrac{q^2}{4M^2})$ are 
\begin{equation*}
\frac{1}{M^2}(p_2+p_4)_{\mu} (p_2+p_4)_{\nu} \big[- G_M \kappa F_2 +
\big(\frac{\kappa F_2}{2 M}\big)^2 (1-q^2/4M^2) \big]  
%\label{eq:E-3}
\end{equation*}
In terms of the Sachs form factors
\begin{equation*}
\big(\frac{\kappa F_2}{2 M}\big)^2 (1-q^2/4M^2) - G_M \kappa F_2 \,=\, 
\frac{G_E^2-G_M^2}{2(1-q^2/4M^2)}
%\label{eq:E-3}
\end{equation*}

\noindent  The square of the matrix element becomes   
\begin{equation}  \begin{split}
 |\mathfrak{M}_{fi}|^2 & ~=~ \frac{4 \pi^2 \alpha^2}{m^2M^2(q^2)^2} 
   \Big[G^2_M (2p_1\cdot p_2 p_3\cdot p_4 +  2p_1\cdot p_4 p_3\cdot p_2
     +(m^2+M^2)q^2)  \\
 & ~~~~ + \frac{G_E^2-G_M^2}{2(1-q^2/4M^2)}(2p_1\cdot(p_2+p_4) p_3\cdot(p_2+p_4)
   +q^2 (p_2+p_4) \cdot(p_2+p_4)/2 ) \Big]  \\
 & ~~=~\frac{4 \pi^2 \alpha^2}{m^2M^2(q^2)^2}  
   \Big[G^2_M (4M^2 EE'+(M^2+m^2)q^2+q^4/2) +  \frac{G_E^2-G_M^2}{1-q^2/4M^2}
   (4M^2 EE'+M^2q^2) \Big] 
\end{split}
\label{eq:matel}
\end{equation}


\begin{thebibliography}{}
\vspace*{-.3cm}       
%\small

\bibitem{experiment}
 A. Afanasev et al.,  Paul Scherrer Institute Proposal R-12-01.1.

\bibitem{rpohl-experiment}
R. Pohl et al.,    %The size of the proton,
Nature 466, 213   

\bibitem{codat06} 
P.J. Mohr, B.N.Taylor, D.B. Newell,  Rev. Mod. Phys. 80, 633 (2008)

\bibitem{mainz2010}
J.C. Bernauer et al., % "High-Precision Determination of the Electric and 
% Magnetic Form Factors of the Proton", 
 Phys. Rev. Lett. 105, 242001 (2010).  An addendum is given in
 arXiv:1108.3533 (2011)  %,  arXiv:1007.5076 

\bibitem{jlab} 
X. Zahn et al., % "High-precision measurement of the proton elastic form
% factor ratio at low $Q^2$
 arXiv:1102.0318 (2011)

\bibitem{RMP} 
E. Borie, G.A. Rinker, %The Energy Levels of Muonic Atoms,
Rev. Mod. Phys. 54, 67 (1982)

\bibitem{Landau}
 V.B. Berestetskii, E.M. Lifshitz, L.P. Pitaevskii, Quantum Electrodynamics,
 Landau-Lifshitz Course on Theoretical Physics Vol.4, Pergamon Press, 
 Oxford, 1982.  (see Eq.(139.4)).

%V.B. Berestetskii, E.M. Lifshitz and L.P. Pitaevskii, 
%Relativistische Quantentheorie, Akademie-Verlag, Berlin, 1975 

%Quantum Electrodynamics, Landau-Lifshitz Course on Theoretical Physics Vol.4,
%2nd edition, Oxford: Butterworth-Heinemann (2007).

\bibitem{Preedom}
B. Preedom, R. Tegen,  %Nucleon electromagnetic form factors from scattering
%of polarized muons or electrons
Phys. Rev. C36, 2466 (1987)

\bibitem{Bjorken-Drell}
J.D. Bjorken, S.D. Drell, {\em Relativistic Quantum Mechanics}, McGraw-Hill,
 New York, 1964.

\bibitem{Tsai}
Y.S. Tsai %
Phys. Rev. 122, 1898 (1961)

\bibitem{Mo-Tsai}
L.W. Mo, Y.S. Tsai %
 Rev. Mod. Phys., 41, 205 (1969)

\bibitem{max-tjon} 
L.C. Maximon, J.A. Tjon, %Radiative corrections to electron-proton scattering, 
 Phys.Rev. C76, 035205 (2007) 

\bibitem{karshenboim}
S.G. Karshenboim,  %What do we actually know about the proton radius?
Can. J. Phys. 77, 241 (1999), Section 4.

\bibitem{kelly} 
J.J. Kelly, %Simple parametrization of nucleon form factors,
 Phys.Rev. C70, 068202 (2004) 

\bibitem{Schwinger}
J. Schwinger %
Phys. Rev. 76, 790 (1949)

\bibitem{Feynman}
R.P. Feynman, %
Phys. Rev. 76, 769 (1949)

\bibitem{akhiezer}
A.I. Akhiezer, V.B. Berestetskii, {\em Quantum electrodynamics}, Wiley
Interscience, New York, 1965.

\bibitem{barbieri}
R. Barbieri, J.A. Micagno, E. Remiddi,  %Electron Form Factors up to   
%Fourth Order.~I,
Il Nuovo Cimento, 11A, 824 (1972)

%\bibitem{jauch}
%J.M. Jauch, F. Rohrlich, {\em The Theory of Photons and Electrons},
%Addison-Wesley, Cambridge, Massachusetts, 1955

\bibitem{barbieri73}
R. Barbieri,  E. Remiddi, % 
Il Nuovo Cimento, 13A, 99 (1973)

\bibitem{hadron1}
E. Borie, %Hadronic Vacuum Polarization Correction in Muonic Atoms,
Z. Phys. A302, 187 (1981)

\bibitem{hadron2}
J.L. Friar, J. Martorell, D.W.L. Sprung, 
%Hadronic vacuum polarization and the Lamb Shift,
Phys. Rev. A59, 4061 (1999) 


%  later add refs 1,2,3 from Max-Tjon (if radiative corrections are included)

\end{thebibliography}
\end{document}